\documentclass[aps,prb,twocolumn,showpacs,superscriptaddress]{revtex4}
\usepackage{graphicx,amsmath,color}

\begin{document}

\title{Stable topological textures in a classical 2D Heisenberg model}

\author{E.~G. Galkina}
\affiliation{Institute of Physics, 03028 Kiev, Ukraine}

\author{E. V. Kirichenko}
\affiliation{Opole University, Institute of Mathematics and
Informatics, Opole, 45-052, Poland}

\author{ B.~A. Ivanov}\email{bivanov@math.uni.opole.pl}
\affiliation{Institute of Magnetism, 03142 Kiev, Ukraine}
\affiliation{National T. Shevchenko University of Kiev, 03127 Kiev,
Ukraine}

\author{V. A. Stephanovich}\email{stef@math.uni.opole.pl}
\homepage{http://cs.uni.opole.pl/~stef}
\affiliation{Opole
University, Institute of Mathematics and Informatics, Opole, 45-052,
Poland}
\begin{abstract}
We show that stable localized topological soliton textures
(skyrmions) with $\pi _2$ topological charge $\nu \geq 1$ exist in a
classical 2D Heisenberg model of a ferromagnet with uniaxial
anisotropy. For this model the soliton exist only if the number of
bound magnons exceeds some threshold value $N_{\rm cr}$ depending on
$\nu $ and the effective anisotropy constant $K_{\rm eff}$. We
define soliton phase diagram as the dependence of threshold energies
and bound magnons number on anisotropy constant. The phase boundary
lines are monotonous for both $\nu=1$ and $\nu
>2$, while the solitons with $\nu=2$ reveal peculiar nonmonotonous
behavior, determining the transition regime from low to high
topological charges. In particular, the soliton energy per
topological charge (topological energy density) achieves a minimum
neither for $\nu=1$ nor high charges, but rather for intermediate
values $\nu=2$ or $\nu=3$.
\end{abstract}
\date{\today }
\pacs{75.10.Hk, 75.30.Ds, 05.45.-a}
\maketitle

\section{Introduction}

The studies of nonlinear excitations of two-dimensional (2D) and
quasi 2D correlated spin systems are an important issue of modern
physics of magnetism, and can be useful for development of general
soliton
concepts.~\cite{one,two,maloz,Bar'yakhtar94,Kosevich90,Bar'yakhtar93}
The topological textures like localized solitons (skyrmions
\cite{Skirm61}) or magnetic vortices make an important contribution
to the thermodynamics of magnetically ordered systems \cite{Huber}
or even determine the character of its ordering as in the case of
Berezinskii-Kosterlitz-Thouless transition.~\cite{Berezinsky72,
Kosterlitz73} Last years the interest for two-dimensional solitons
has grown since they are frequently realized as ground state in the
finite-size mesoscopic magnetic samples, so-called magnetic
dots.~\cite {skomski}

The most ``famous'' topological solitons are magnetic vortices
having $\pi _1$ topological charge. These vortices are usually
related to thermodynamic aspects of soliton
physics.~\cite{Berezinsky72, Kosterlitz73} Also, they appear in
mesoscopic nanostructures.~\cite {skomski} Easy-plane magnets with
continuously degenerated ground state have vortices with the energy
being logarithmically divergent as a function of system size. The
other and much less studied example of topological solitons are
magnetic skyrmions which are present in isotropic or easy-axis
magnets. Contrary to the above vortices, the latter textures are
characterized by nontrivial $\pi _2$ topological charge and finite
energy. It is known that they determine the response functions of 2D
magnets at finite temperatures \cite{thermodyn1,thermodyn2} and take
part in long-range order breakdown in isotropic
magnets.~\cite{Belavin75} The skyrmions form ground state of
magnetic nanoparticles with easy-axis anisotropy.~\cite{nikos} Their
analysis is more complicated as compared to magnetic vortices and
comprises many nontrivial features. An important example of latter
features is the problem of a skyrmion stability since due to
Hobart-Derrick theorem the static solitons with finite energy are
unstable for wide class of models including standard continuous
magnetic models.~\cite{hob,der}

For magnetic vortices, the consideration of lowest possible
topological charge $\nu =1$ is sufficiently as the vortex energy
grows with $\nu $, $E_{\nu }^{\rm{vort}}\propto \nu ^2$. Because of
that it is advantageous for a vortex with $\nu >1$ to decay for
$\nu$ vortices with $\nu =1$ and the vortices with $\nu=2$ can be
stable in exceptional cases only.~\cite{Wysin05} The situation for
skyrmions is not that simple. The simplest continuous model for
isotropic 2D ferromagnet (FM)
\begin{equation}\label{qe1}
W^{\rm{is}}=\frac{Ja^2S^2}{2}\int [\nabla {\vec m}]^2d^2x
\end{equation}
admits the well-known Belvin-Polakov (BP) solution,~\cite{Belavin75}
which reads
\begin{equation}\label{qe2}
\tan \frac{\theta }{2} =\left(\frac{R}{r}\right)^\nu,\
\varphi=\varphi _0+ \nu \chi,
\end{equation}
where $\vec{m}$ is normalized magnetization
\begin{equation}\label{gan}
\vec{m} =
\left(\sin\theta\cos\varphi;\sin\theta\sin\varphi;\cos\theta\right),
\end{equation}
$S$ is a spin value, $a$ is a lattice constant of a 2D FM, $J$ is
its exchange constant, $r$ and $\chi$ are polar coordinates in the
$XY$--plane, $\varphi_0$ is an arbitrary constant. Solution
\eqref{qe2} has the energy
\begin{equation}
E_{0\nu }=\nu E_0, \ E_0=4\pi JS^2 \label{ebp0}
\end{equation}
so that the state of BP skyrmions with $\nu >1$ merges or
dissociates into several other similar skyrmions with different $\nu
$'s (the only rule that in such process the topological charge
should conserve) without their energy altering. Such exact
degeneration is related to very high hidden symmetry, stemming from
exact integrability of corresponding static model \eqref{qe1} (see,
e.g.,~\cite{Belavin75}). This degeneration should certainly be
lifted if we go beyond the model \eqref{qe1}. The most important
characteristic here is the parameter ${\mathcal E}_\nu$, which is
appropriate to call the {\em {topological energy density}},
\begin{equation}\label{qe3}
{\mathcal E}_\nu=\frac{E_\nu}{\nu},
\end{equation}
where $E_\nu$ is the energy of a soliton with topological charge
$\nu$. If ${\mathcal E}_\nu$ is a growing function of $\nu$, the
most favorable state with a given $\nu$ comprises $\nu $ solitons
with unit topological charge otherwise such state is unstable.
Latter question is especially important for the investigation of
general regularities of the highly excited magnet states evolution,
(see, e.g. Ref.~\onlinecite{dziar} and references therein) or for
the analysis of essentially inhomogeneous magnet states under strong
pumping.~\cite{lvov} Latter states can be generated by the ultrafast
pulses of magnetic field, see Refs.~\onlinecite{kim1,kim2} for
details. The preceding discussion demonstrates that the problem of
obtaining and investigation of the stable skyrmions with higher
topological charges  is extremely important.~\cite{manton}

The present work is devoted to the analysis of skyrmions with higher
$\pi _2$ topological charges in 2D Heisenberg ferromagnet with
uniaxial anisotropy \eqref{eq:H-discrete}. We show that there exists
a certain range of system parameters (exchange and anisotropy
constants), where stable precessional solitons with topological
charge  $\nu>1$ exist. It turns out that in wide range of anisotropy
constants, the topological energy density ${\cal E}_\nu$ of the
textures with $\nu>1$ is lower then that of the textures with $\nu
=1$. On the other hand, the solitons with $\nu=1$ and $\nu >2$ have
monotonously growing phase boundary functions ${\cal E}_{\nu , {\rm
cr}}(N_{\nu , {\rm cr}})$, while the case $\nu=2$ has peculiar
nonmonotonous behavior, determining the transition regime from low
to high topological charges. This means that the preferable values
of soliton topological charge are neither $\nu=1$ nor high charges,
but rather $\nu=2$ or $\nu=3$.

\section{Model description and soliton classification}

We begin with the discrete model of a classical 2D FM with uniaxial
anisotropy, described by the following Hamiltonian
\begin{eqnarray} \label{eq:H-discrete}
&&{\mathcal H} = -\frac 12 \sum_{{\vec n},{\vec a}}\!
\left(J\vec{S}_{\vec n}\!\cdot\! \vec{S}_{{\vec n}+{\vec a}}
+\kappa S^z_{\vec n} S^z_{{\vec n}+{\vec a}} \right)+ \nonumber \\
&&+K\sum_{\vec n} [(S^x_{\vec n})^2+(S^y_{\vec n})^2].
\end{eqnarray}
Here  $\vec{S}\equiv\left(S^x, S^y, S^z\right)$ is a classical spin
vector with fixed length $S$ on the site $\vec{n}$ of a 2D square
lattice. The summations run over all lattice sites $\vec{n}$ and
nearest--neighbors ${\vec a}$, $J > 0$ is the exchange integral and
the constant $\kappa $ describes the anisotropy of spin interaction.
In subsequent discussion, we refer to this type of anisotropy as
exchange anisotropy (EA). Additionally, we took into account
single-ion anisotropy (SIA) with constant $K$. We consider $z$--axis
to be easy magnetization direction so that $K>0$ or $\kappa >0$.

The analysis of real magnetic systems with discreet spins can be
performed only numerically. In principle, it can be done by the same
method as was described in Ref.~\onlinecite{my}, but it is not easy
to extract necessary information from the set of numerical data. On
the other hand, if we neglect the specific effects of discreteness
(which appear at strong anisotropy only, $K,\kappa \geq J$), like
the presence of pure collinear structures, see Ref.~\onlinecite{my},
the consideration can be simplified using the generalized continuous
approximation and classical Landau-Lifshitz equation. In this
approximation we can introduce the smooth function $\vec S(x,y,t)$
instead of discreet variable $\vec S_n(t)$. In this case, the
classical magnetic energy functional $W[\vec{S}]$ can be constructed
expanding the discrete Hamiltonian \eqref{eq:H-discrete} in power
series of magnetization gradients, yielding
\begin{equation}\label{kl1}
    E_\nu=W_2+W_4+...,
\end{equation}
where $W_2$ contains zeroth and second order contributions to
magnetic energy, see \eqref{qe1}, and $W_4$ contains the fourth
powers of gradients. The explicit (and quite cumbersome) expressions
for $W_2$ and $W_4$ had been presented in Ref.~\onlinecite{my}. We
note here, that single-ion anisotropy enters only $W_2$, but not
$W_4$ and higher terms, while exchange anisotropy enters every term
of the expansion \eqref{kl1}.

For isotropic case $K=0$ and $\kappa =0$, $W_2$ coincides with the
energy of the isotropic continuous model \eqref{qe1} bearing BP
soliton solutions of the form \eqref{qe2} with degenerate (with
respect to topological charge) topological energy density
\eqref{qe3}, see also \eqref{ebp0}. A simple accounting of magnetic
anisotropy in $W_2$ generates a model, which is typical example of
the models governed by Hobbard-Derrick theorem - it does not admit
static stable soliton solutions. In the model with $W=W_2$ only, the
size of any texture like domain wall, soliton etc (see, e.g. Ref.
\onlinecite{my} for details) is given by the characteristic length
$l_0$
\begin{equation}\label{lzero}
l_0^2=\frac{a^2J}{2K_{\rm eff}},\ K_{\rm eff}=K+2\kappa.
\end{equation}
In the case of weak anisotropy, $K_{\rm eff} \ll J$, the length
scale $l_0 \gg a$ so that the magnetization varies slowly in a
space.

Now we consider generalized model \eqref{kl1}, including higher
powers of gradients. In the expansion \eqref{kl1}, we limit
ourselves to the terms of fourth order only as they are playing a
decisive role in solitons stabilization, see
Refs.~\onlinecite{Ivanov86,IvWysin02} for details. The simplest
possible generalized model with account for $W_4$ only can in
principle admit the above static stable solitons.~\cite{two} Simple
scaling arguments can demonstrate that. Namely, if the soliton core
has size $R$, the specific calculations with energy functionals like
\eqref{asm7} (see below) yield the dependence of the energy $E_\nu $
on $R$ in the form
\begin{equation}\label{kl2}
    E_\nu=-A_{\nu}\frac{ l_0^2}{R^2}+B_\nu+C_\nu \frac{R^2}{l_0^2} ,
\end{equation}
where the first term comes from $W_4$ and the rest come from $W_2$.
If $E_\nu (R)$ had a minimum (this occurs if $A_\nu <0$), this would
mean the existence of a stable static skyrmion. Unfortunately, for
the real model of magnet \eqref{eq:H-discrete} $A_\nu
>0$ so that the first term is negative and the dependence $E_\nu
(R)$ does not have a minimum. We note also that the expansion
\eqref{kl1} and the above scale arguments are valid for any symmetry
of initial 2D discrete lattice. The only difference is in the
coefficients before gradient powers. Although these coefficients
influence the solitons properties, the main feature of these
expansions, consisting in the fact that in Eq. \eqref{kl2} $A_\nu>0$
remains the same. In other words, we did not find any symmetry of 2D
FM with nearest-neighbor ferromagnetic interaction, where $E_\nu(R)$
has a minimum so that stable static soliton can exist.~\cite{sn}

In the absence of \emph{static} two--dimensional  solitons, it is
possible to construct stable soliton states with \emph{stationary
dynamics}, due to the presence of additional integrals of motion for
magnetization fields. The purely uniaxial model
\eqref{eq:H-discrete} possesses the exact symmetry with respect to
the spin rotation around $z-$axis so that the energy functional
$E_{\nu }[\theta,\varphi]$ does not depend explicitly on the
variable $\varphi$.  This leads to the appearance of  an additional
integral of motion, $z$-projection of total spin. This integral of
motion can be conveniently parameterized via integer $N$ defining a
number of bound magnons in a soliton $N$, see
Ref.~\onlinecite{Kosevich90} for details. In continuous
approximation it can be written as
\begin{equation}\label{nen}
N = \frac{S}{a^2}\int d^2 x \left(1-\cos\theta\right).
\end{equation}
Conservation of $N$ leads to the presence of so--called
\emph{precessional} solitons characterizing by time-independent
projection of magnetization onto the easy $z-$axis and with the
precession of the magnetization vector $\vec m$ at constant
frequency $\omega$ around the $z$ axis,
\begin{equation}\label{omega}
\theta = \theta (r), \, \varphi=\omega t+\nu \chi+\varphi _0,
\end{equation}
which holds instead of (\ref{qe2}) in this case. The analogs of such
precessional solitons are known to occur in different
field-theoretical models, it is enough to note the non-topological
Coleman's $Q$-balls,~\cite{colemanQ} which do not have topological
properties as well as  $\pi _2$ topological $Q$ -
lumps,~\cite{Qlumps} see Ref.~\onlinecite{two} for details.

Stable dynamical solutions with nonzero $\omega$ correspond to
conditional  (for fixed $N$ value) minimum of the energy functional
$E_{\nu }$. Namely, we may look for an extremum of the expression
\begin{equation}\label{cond}
L=E_\nu-\hbar \omega N,
\end{equation}
where $\omega $ is an internal soliton precession frequency, which
in this case can be regarded as Lagrange multiplier. Note, that
functional \eqref{cond} is nothing but the Lagrangian of 2D FM
magnetization field calculated with respect to specific time
dependence \eqref{omega}. This condition leads to the relation
\cite{Kosevich90}
%\begin{equation}\label{dEdN}
$\hbar \omega= dE_\nu /dN$,
%\end{equation}
which determines the microscopic origin of the precessional
frequency $\omega$. Namely, an addition of one extra spin deviation
(bound magnon) to a soliton changes its energy by $\hbar \omega$.
Thus, the dependencies $E_\nu (N)$ and $\omega (N)$ are very
important for the problem of a soliton stability.

\section{The structure and stability of skyrmions. Critical energy.}
Further analysis of above continuous model consists in the solution
of differential equations for soliton structure, what is equivalent
to the minimization of the functional \eqref{kl1}. As these
equations can barely be solved analytically, here we analyze the
solitons properties by direct variational method. As we have shown
earlier by comparison of variational approach and direct numerical
minimization of initial discreet energy on a lattice,~\cite{my} the
variational approach gives fairly good results for weak anisotropies
$K_{\rm eff} \leq 0.5 J$, where the generalized continuous
description is valid.

The continuous models like \eqref{qe1} and \eqref{kl1} are usually
parameterized by angular variables \eqref{gan} so that the energy
$E_\nu$ becomes a functional of these variables, $E_\nu \equiv E_\nu
[\theta, \nabla \theta, \phi, \nabla \phi]$. Having energy
functional $E_\nu$, we can write the corresponding Landau-Lifshitz
equations and Lagrangian \eqref{cond}.

To apply direct variational method for minimization of the energy
$E_\nu=W_2+W_4$ we use the trial function
\begin{equation}\label{macd}
\tan \frac{\theta }{2}=\frac{2^{1-\nu}(\Lambda R)^\nu}{(\nu-1)!}
K_\nu\left( \Lambda r\right),
\end{equation}
where $K_\nu (x)$ is the McDonald function with index $\nu $.
~\cite{grad} Note, that trial function (\ref{macd}) is based on the
interpolative solution, constructed in Ref.~\onlinecite{Voronov83}.
The trial function \eqref{macd} gives correct asymptotics both for
$r\to 0$ (corresponding to BP soliton) and for $r\to\infty$
(exponential decay with some characteristic scale $1/\Lambda $), see
Refs.~\onlinecite{Voronov83,Ivanov86,Kosevich90} for details. Latter
exponential asymptotics is absent for solitons in isotropic 2D FM,
e.g. for BP soliton \eqref{qe2}. It can be shown that the
exponential asymptotics occurs for anisotropic models, where the
length of decay is proportional to $l_0$ \eqref{lzero}. It had been
demonstrated  that due to power-law asymptotics of $\theta (r)$ for
isotropic magnet, the integrals defining soliton energy and number
of bound magnons (see below Eqs. \eqref{asm7}) are divergent for
$\nu=1$.~\cite{Voronov83,Ivanov86,Kosevich90} To avoid this
divergency in anisotropic models like \eqref{eq:H-discrete}, the
interpolative (between BP asymptotics at $r\to 0$ and exponential
one at $r\to \infty$) solution had been put forward in Ref.
\onlinecite{Voronov83}. Our analysis of continuous model \eqref{kl1}
shows that at $\nu=1$ the above divergence can be cut off by the
exponential asymptotics only. For $\nu=2$ the power-law asymptotics
is sufficient for convergence of corresponding integrals, while
their derivatives with respect to  parameter $\Lambda$ are
divergent. In this case, the divergences in derivatives are also
eliminated by exponential asymptotics. At $\nu>2$ all integrals and
their derivatives (with respect to $\Lambda$) are convergent. This
means that the behavior of solitons with $\nu=1$ and 2 on one side
and those with $\nu>2$ on the other side is different, being
determined by the interplay between effects of anisotropy and higher
spatial derivatives. %This fact is corroborated by the
%Ref.~\onlinecite{Wysin05}, where it was shown that stable localized
%topological textures with $\nu>1$ can exist in a 2D magnet if the
%physical vacancies (missing spins in its lattice sites) are present.

In our minimization method, the parameter $\Lambda $ is variational,
while the parameter $R$ is kept constant as it is related to $N$,
$N\propto R^2$, see, e.g. \cite{Kosevich90}. In other words, we
minimize the energy $E$ with the trial function (\ref{macd}) over
$\Lambda $ for constant $R$. This approach has the advantage that it
also permits to investigate the stability of obtained soliton
texture. Namely, a soliton is stable if it corresponds to the
conditional minimum of the energy at fixed $N$, and it is unstable
otherwise.

To proceed further, we introduce following dimensionless variables
\begin{equation}\label{dimvar}
x=\Lambda r, \  \lambda=a\Lambda,\ z= \Lambda R.
\end{equation}
and express trial function \eqref{macd} in terms of them. We have
\begin{equation}\label{trf}
\tan \frac{\theta }{2} =\frac{2^{1-\nu}z^\nu}{(\nu -1)!}\ K_\nu(x).
\end{equation}
Then using the equation \eqref{nen}, we can calculate the number of
bound magnons in the soliton
\begin{equation}\label{n1}
\frac{N}{S}=\frac{2\pi}{\lambda ^2} \int_0^{\infty}(1-\cos
\theta)xdx\equiv \frac{2\pi}{\lambda ^2}\psi(z).
\end{equation}
In the variables \eqref{dimvar} the soliton energy assumes the form
\begin{equation}\label{dimen}
  \frac{{\mathcal E}_\nu}{2\pi JS^2}= \frac{K_{\rm{eff}}}{\lambda
^2}\gamma _0(z)+\gamma _2(z)-\frac{1}{24}\lambda ^2\gamma _4(z),
\end{equation}
where
\begin{widetext}
\begin{eqnarray}
&&\gamma _0(z)=\int_{0}^{\infty }\sin ^{2}\theta xdx, \quad \gamma
_2(z)=\frac{1}{2}\int_0^{\infty }xdx\left[ \theta^{\prime 2}
\left( 1+\kappa \sin ^2\theta \right) +\frac{%
\nu ^2\sin ^2\theta }{x^2}\right] ,  \nonumber \\
&&\gamma _4(z)=\int_0^{\infty }xdx\Biggl\{\left( \Delta _x\theta
\right)^2\left( 1+\kappa \sin ^2\theta \right) +\theta^{\prime 4}
\left( 1+\kappa \cos ^2\theta \right) +\frac{\nu^2\sin ^2\theta }{x^2}%
\left( \frac{\nu^2}{x^{2}}+2\theta ^{\prime 2}\right) +  \nonumber \\
&&+\Delta _{x}\theta \sin 2\theta \left( \kappa \theta ^{\prime 2}-\frac{%
\nu ^2}{x^2}\right) \Biggr\},\ \ \theta '=\frac{d\theta}{dx}, \
\Delta _x\theta = \frac{d ^2 \theta}{dx^2} +\frac 1x\frac{d
\theta}{dx}\;. \label{asm7}
\end{eqnarray}
\end{widetext}
Thus, we express the energy and the number of magnons via two
parameters, $\lambda$ and $z$. It turns out, that initial
dimensional variables $\Lambda $ and $R$ enter the problem only in
the form of their product $z$. The dependence of $N$ and $E_\nu$ on
$z$ enters the problem via a few complicated functions $\psi,
\gamma_0 , \gamma_2 $ and $\gamma_4$, which can be written only
implicitly in the form of integrals \eqref{asm7}. However, in terms
of these functions, the dependence on $\lambda $ (\ref{dimen}) turns
out to be quite simple. This permits reformulation of the initial
variational problem in terms of variables $z$ and $N$ only. Namely,
we express
\begin{equation}\label{tyi}
  \lambda ^2=\frac{2\pi}{(N/S)}\psi(z)
  %\int_0^{\infty}(1-\cos
%\theta)xdx
\end{equation}
and substitute this expression into the dimensionless energy
(\ref{dimen}). This gives us the expression for the energy of a
soliton with given $N$, as a function of variational parameter $z$.
Then we can find a minimum of $E_\nu$ with respect to $z$, keeping
$N$ constant.

\begin{figure}
\vspace*{-5mm} \hspace*{-5mm} \centering{\
\includegraphics[width=0.5\textwidth]{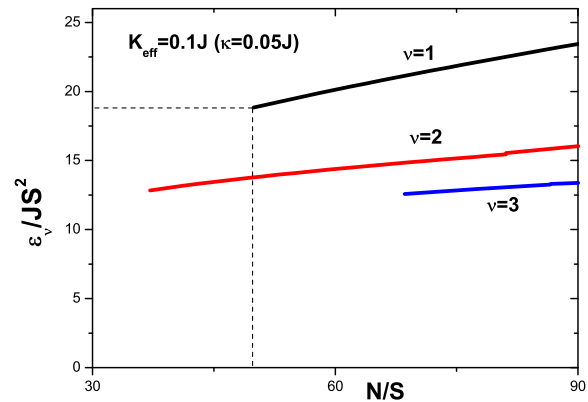}}
\caption{Dependence of a soliton topological energy density
${\mathcal E}_{\nu}(N)$ for skyrmions with different $\nu$'s and
exchange anisotropy only, $\kappa=0.05J$ ($K_{\rm {eff}}=0.1J$).
Dashed lines show ${\mathcal E}_{\nu , \rm{cr}}(\nu)$ and $N_{\nu ,
\rm{cr}}$ for $\nu = 1$.}
%Inset - the same plotted versus
%$N/N_{\rm cr}(\nu)$ showing that at $\kappa=0.05J$ the energy of
%solitons with $\nu>1$ are always smaller then that for $\nu=1$.}
\label{fig:v1}
\end{figure}
%\vspace*{-5mm}

The result of such numerical minimization in the form of the
dependence ${\mathcal E}_\nu (N)$ is shown in Fig.\ref{fig:v1} for a
magnet with purely exchange anisotropy $\kappa=0.05$. The curves for
higher $\kappa $ are qualitatively the same. To justify the
applicability of our direct variational approach, earlier we have
shown,\cite{my}  that for $\nu=1$ the dependencies ${\mathcal E}_\nu
(N)$ found by variational and numerical minimizations of the energy,
are identical at small enough anisotropy, $K_{\rm{eff}} \leq 0.5J$.

Our analysis shows (see also Fig.\ref{fig:v1}) that topological
energy density grows slowly as function of $N$. We note here that
this property holds for all anisotropy constants and topological
charges. Even more interesting is the fact that all curves
${\mathcal E}_\nu (N)$ have threshold points ${\mathcal E}_{\nu ,
\rm {cr}}={\mathcal E}_\nu (N_{\nu , \rm cr})$ so that solitons
exist only at ${\mathcal E}_{\nu}>{\mathcal E}_{\nu , \rm {cr}}$ and
$N_{\nu}>N_{\nu , \rm {cr}}$. These threshold values determine the
minimal soliton energy, which is most important characteristic of
soliton contribution into magnet thermodynamics and can be observed
experimentally.~\cite{thermodyn1} For instance (see
Fig.\ref{fig:v1}), at $\nu =1$ and $\kappa=0.05J$ ${\mathcal E}_{\nu
, \rm {cr}}\approx 18 JS^2$, which is substantially higher then that
expected from Eq. \eqref{ebp0} $4\pi JS^2\approx 12.56 JS^2$.

Fig. \ref{fig:v1} demonstrates one more unexpected soliton property,
namely that at $\kappa=0.05J$ the energy density ${\mathcal E}_\nu
(N)$ {\it decreases} with increase of $\nu$ and fixed $N$ value. As
we will show below, such behavior occurs for many values of $K_{\rm
{eff}}$, although there can be exceptions. The most important
feature, however, is the existence of above threshold energy and
bound magnons number values. The behavior of these values at
variation of anisotropy constants is quite nontrivial. For example,
at $\kappa=0.05J$ (the value chosen for Fig. \ref{fig:v1}) $N_{\rm
cr}(\nu=1)> N_{\rm cr}(\nu=2)$ but $N_{\rm cr}(\nu=3)> N_{\rm
cr}(\nu=1)$. At the same time, the corresponding threshold energies
behave monotonically ${\mathcal E}_{\rm {cr}}(\nu=1)>{\mathcal
E}_{\rm {cr}}(\nu=2)>{\mathcal E}_{\rm {cr}}(\nu=3)$. Our extensive
analysis of numerical curves ${\mathcal E}_\nu (N)$ for different
anisotropies have shown that  their overall behavior is dictated
primarily by the threshold values: if ${\mathcal E}_{\nu , \rm
{cr}}>{\mathcal E}_{\nu ' , \rm {cr}}$, then the entire curve
${\mathcal E}_\nu (N)$ lies above corresponding curve ${\mathcal
E}_{\nu '} (N)$ in wide interval of $N$'s. This shows the importance
of the above threshold values for the properties of solitons. The
information about these values can be conveniently represented in
the form of so-called solitons phase diagram, i.e. the dependence of
${\mathcal E}_{\rm cr}$ and $N_{\rm {cr}}$ on $K_{\mathrm{eff}}$.

\section{Solitons phase diagram.}
To obtain the phase diagram, we should pay attention to the details
of above minimization procedure. Namely, to obtain truly stable
soliton texture we should demand that the conditional extremum of
the energy $E_\nu $ is a minimum. So, we keep track not only to the
first derivative $dE_\nu/dz$ to be zero but also to the second
derivative to be positive (corresponding to a minimum) at the point
$z_{\rm {min}}$, where $dE_\nu/dz=0$. When a soliton approaches the
limit of its stability, the modulus of second derivative diminishes,
becoming zero at the stability limit $z_{\rm cr}$, corresponding to
above values ${\mathcal E}_\nu \equiv {\mathcal E}_{\nu , \rm {cr}}$
and $N=N_{\nu , \rm {cr}}$. Thus the above soliton phase diagram is
indeed determined by the instability point $z_{\rm cr}$ (as a
function of ratio $K_{\rm{eff}}/J$) where both first and second
derivatives of the energy are zero.
\begin{figure}
\vspace*{-5mm} \hspace*{-5mm} \centering{\
\includegraphics[width=80mm]{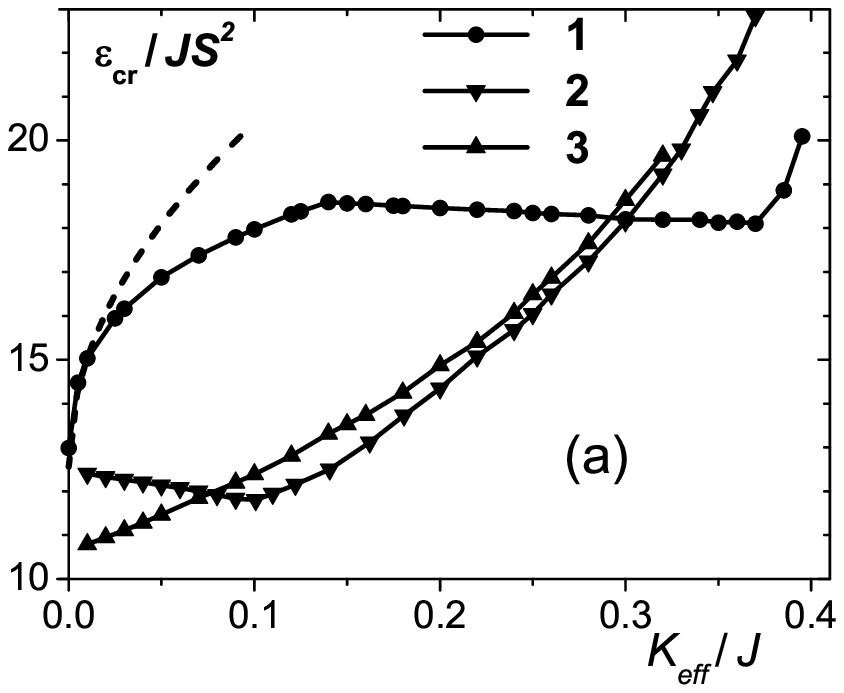}
\includegraphics[width=80mm]{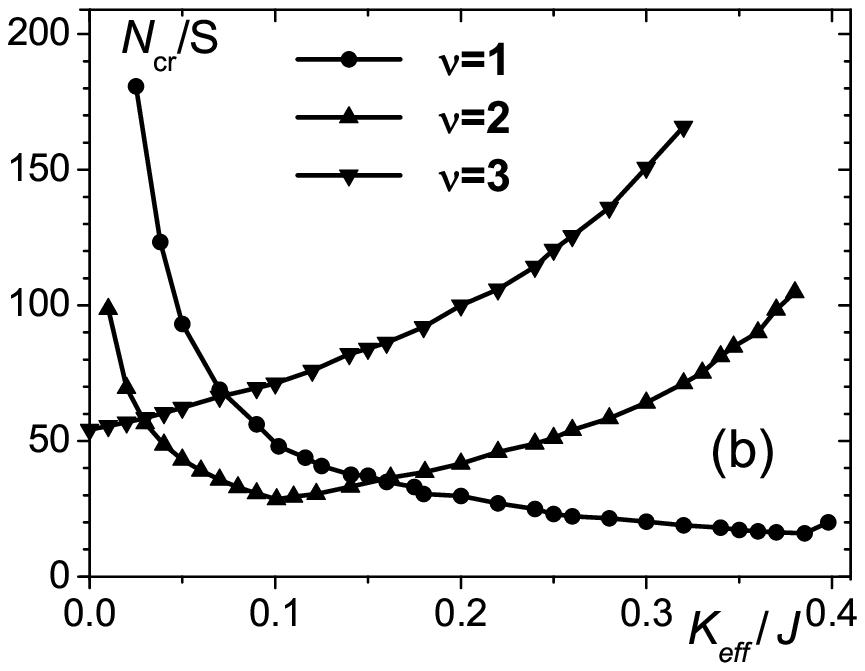}}
\caption{Soliton phase diagram. (a) - threshold topological energy
density (dashed line at small $K_{\rm eff}$ is described by
Eq.\eqref{ass12a}), (b) - threshold magnon number. Solitons exist at
${\mathcal E}>{\mathcal E}_{\rm cr}$ and $N>N_{\rm cr}$. For small
$\kappa $ the value of $N_{\rm {cr}}$ is divergent as
$1/\sqrt{K_{\rm eff}}$ at $\nu=1$ and 2 and is constant at $\nu \geq
3$.} \label{fig:v2}
\end{figure}

The shape of the above phase diagram is determined by the character
of anisotropy and $\nu $ value and is reported on Fig.~\ref{fig:v2}
for exchange anisotropy. First of all, one can see that the
dependence ${\mathcal E}_{\rm cr}(\nu )$ is quite complicated and
its character change twice, at  $K_{\rm{eff}} \approx 0.07$ and at
$K_{\rm{eff}} \approx 0.29$. It is seen from Fig.\ref{fig:v2}a that
while there is a quite large region of $K_{\rm eff}$ values, where
${\mathcal E}_2$ and ${\mathcal E}_3$ are smaller then ${\mathcal
E}_1$, the region where ${\mathcal E}_3<{\mathcal E}_2$ is
approximately three times smaller. We have shown that ${\mathcal
E}_4$ is always larger then ${\mathcal E}_3$ although still smaller
then ${\mathcal E}_1$. The energy ${\mathcal E}_{\rm cr}(\nu )$ is
growing with $\nu $  at $K_{\rm eff}\geq 0.29J$  only.

Fig.~\ref{fig:v2}b demonstrates the divergence of $N_{\rm{cr}}$ at
small $K_{\rm{eff}} $ as $\nu=1$ and 2. This is related to the fact
that in BP soliton the integral describing $N$ diverges
logarithmically as $r\to \infty$ for $\nu =1$, see also above.

\begin{figure}
\vspace*{-5mm} \hspace*{-5mm} \centering{\
\includegraphics[width=80mm]{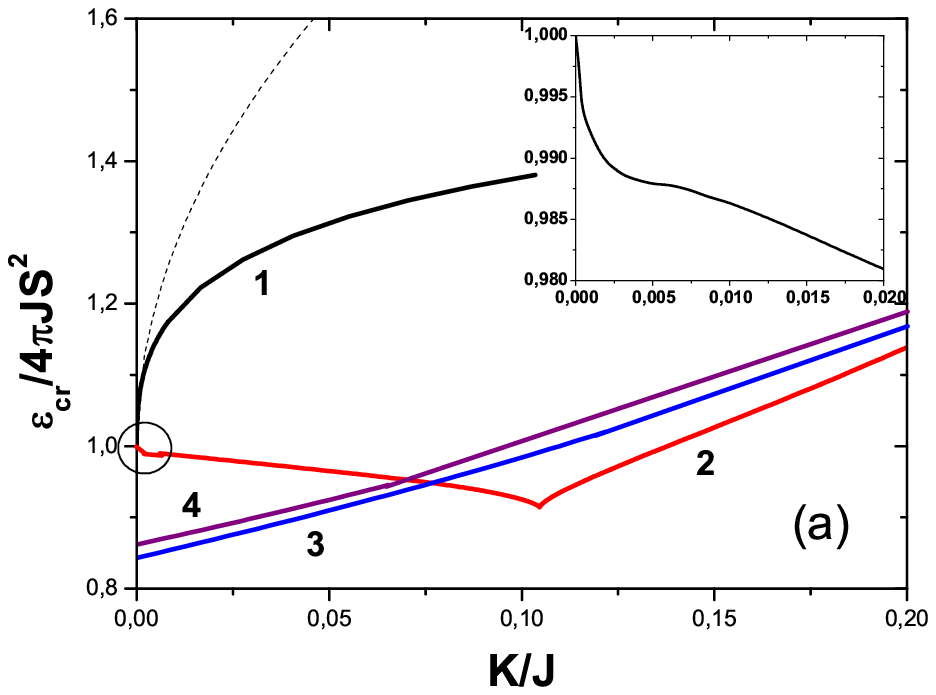}
\includegraphics[width=80mm]{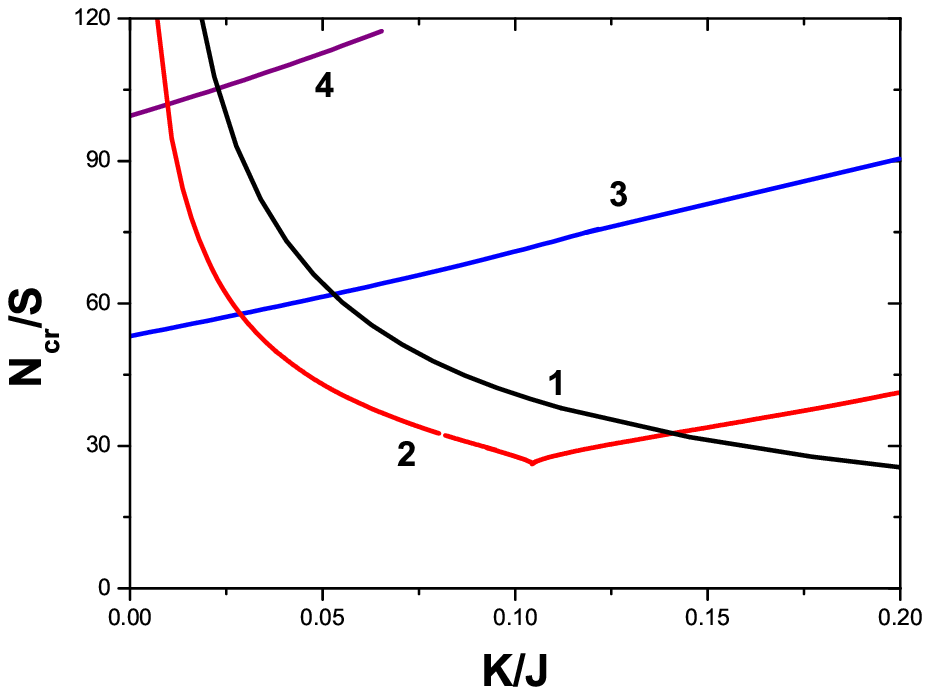}}
\caption{Soliton phase diagram for the model with uniaxial
anisotropy only. (a) - threshold topological energy density (dashed
line at small $K/J$ is asymptotics \eqref{ass1}). Inset expands the
encircled area on the main panel and shows the behavior of
${\mathcal E}_{2, \rm cr }$ at small $K/J$. (b) - threshold magnons
number. The ranges where solitons exist are similar to those on
Fig.~\ref{fig:v2}. Numbers near curves correspond to $\nu$ values.}
\label{fig:w4}
\end{figure}

Overall, Fig.~\ref{fig:v2} demonstrates the complicated behavior of
solitons phase diagram, which is needed to be understood. To
understand better the above complex behavior, it is instructive to
obtain the phase diagram analytically.  Without loss of generality,
we shall do so for the case of uniaxial anisotropy only. Such
analytical treatment is possible since the anisotropy constant $K$
in this case enters the problem only via coefficient in the first
term of Eq. \eqref{dimen}. This permits to obtain the analytic
dependencies ${\mathcal E}_{\nu , \rm {cr}}(K)$ and $N_{\nu , \rm
{cr}}(K)$ in the inverse form $K({\mathcal E}, N)$. Note, that the
structure of Eq. \eqref{asm7} suggests that the above analytical
procedure is applicable for both exchange and uniaxial anisotropy.
Our analysis shows that this does not change the situation
qualitatively. Moreover, the quantitative results are close to each
other, compare Figs.~\ref{fig:v2} and \ref{fig:w4}.

As it was shown above, the soliton phase diagram is determined by
the instability point $z_{\rm cr}$. To obtain an equation for latter
point, we rewrite the Eq. \eqref{dimen} in the form
\begin{eqnarray}\label{ey1}
&&\frac{{\mathcal E}_\nu}{2\pi JS^2} = a(z)y + b(z) - \frac{c(z)}{y},\\
&&y=\frac{N}{S},\ a(z)=\frac{K\gamma_0(z)}{2\pi J \psi(z)},\nonumber \\
&&b(z)=\gamma_2(z),\ c(z)=\frac{\pi}{12}\gamma_4(z)\psi(z) \nonumber
\end{eqnarray}
(function $\psi(z)$ is defined by Eq. \eqref{n1}) and equate to zero
the first and second derivatives of ${\mathcal E}_\nu$ with respect
to $z$. This gives the equation for the dependence $y(z_{\rm cr})$
in the form
\begin{equation}\label{ey2}
y_{\rm cr}\equiv y(z_{\rm cr}) =  - \frac{b'(z_{\rm cr})c''(z_{\rm
cr}) - c'(z_{\rm cr})b''(z_{\rm cr})}{a'(z_{\rm cr})c''(z_{\rm cr})
- c'(z_{\rm cr})a''(z_{\rm cr})},
\end{equation}
where primes mean corresponding derivatives. Then, the equation for
$z_{\rm cr}$ can be obtained by substitution of above $y_{\rm cr}$
into the one of the equations determining zero of first or second
energy derivatives with respect to $z$. It turns out that this
equation always has real solutions if we formally admit the
existence of negative $K$.

The dependence \eqref{ey2} permits to obtain the equation for the
phase diagram in the implicit form $K(z_{\rm{cr}})$, which reads
\begin{eqnarray}\label{ey3}
&&K(z_{\rm{cr}}) = \frac{J}{{c''(z_{\rm{cr}})}}\left[ {\frac{{q
''(z_{\rm{cr}})}}{\pi }y_{\rm cr}^2  + b''(z_{\rm{cr}})y_{\rm cr} }
\right],\nonumber \\
&&q(z)=\frac{\gamma_0(z)}{2\psi(z)}.
\end{eqnarray}

\begin{figure}
\vspace*{-5mm} \hspace*{-5mm} \centering{\
\includegraphics[width=0.5\textwidth]{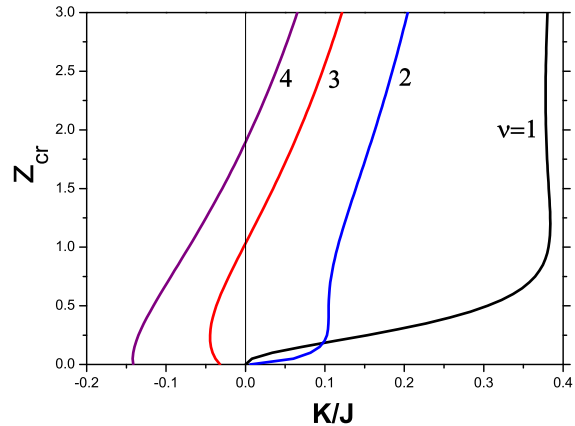}}
\caption{Dependence $z_{\rm{cr}}(K/J)$. Numbers near the curves
correspond to $\nu $ values.} \label{fig:w3a}
\end{figure}

The dependence $z_{\rm{cr}}(K)$, obtained by inversion of Eq.
\eqref{ey3}, is reported on Fig.~\ref{fig:w3a}. It is seen that
while the entire curves $z_{\rm{cr}}(K)$ at $\nu=1$ and 2 lie at
$K>0$, the curves for $\nu>2$ lie in this range only partially.
Since for uniaxial anisotropy $K$ can be only positive, only those
parts of the curves with $\nu>2$ where $K>0$ correspond to
physically realizable case. This clarifies the reason why the
dependence $N_{\rm{cr}}(K/J)$ for skyrmions with $\nu>2$ begins from
finite $N$ values, see Fig. \ref{fig:v2}b. This different behavior
is related to the different kind of convergence of corresponding
integrals at $\nu=1$ and 2 as well as for $\nu>2$, see discussion
above.

The dependence $z_{\rm{cr}}(K/J)$ can be easily recalculated to the
solitons phase diagram ${\mathcal E}_{{\rm cr}\nu}(K/J)$ and
$N_{{\rm cr}}(K/J)$. This diagram is shown on Figs. \ref{fig:w4}a,b.
It is seen the qualitative coincidence with the numerical curves
from Figs.\ref{fig:v2}a,b. The details of behavior can now be better
seen then from above numerical curves.

Our analysis shows that aforementioned different character of
convergence of integrals influences asymptotics of the phase diagram
curves at small anisotropies. This influence can be seen from Figs.
\ref{fig:v2}a and \ref{fig:w4}, where the dependence ${\mathcal
E}_{\nu , \rm {cr}}(K/J)$ at $\nu=1$ and 2 is nonanalytical at
$(K/J)\to 0$ , while for $\nu>2$ it is analytical. The reason for
such behavior can be seen from Fig. \ref{fig:w3a}, showing that
small $K/J$ correspond to small $z$ for $\nu=1$ and 2, while for
$\nu >2$ all values of $K/J$ including limiting case $K/J \to 0$
correspond to finite $z$. This means that at $\nu
>2$ the asymptotic analysis of the curves ${\mathcal E}_{\nu , \rm {cr}}(K/J)$
can be done simply by Taylor expansion at small $K/J$,
which yields simple monotonic behavior of ${\mathcal E}_{\nu , \rm
{cr}}$ and $N_{\nu , \rm {cr}}$.

At $\nu =1$ and 2 the situation is not so simple and requires more
complicated analysis. Such analysis can be performed on the base of
Eqs. \eqref{ey2}, \eqref{ey3} and is quite cumbersome. The main idea
is that functions $a(z)$, $b(z)$ and $c(z)$ in \eqref{ey1} at small
$z$ can be represented via exponential integral
functions,~\cite{grad} which, in turn, may be expanded in asymptotic
series. This procedure for $\nu =1$ gives following parametric
dependence
\begin{eqnarray}
&&\frac{K}{J}=\frac{{18}}{{\ln 3}}z^4 \ln ^6 z,\  \frac{{\mathcal
E}_{\rm cr
}}{4\pi JS^2 }=1 + \frac{4}{{\ln 3}}z^2 \ln ^4 z,\nonumber \\
&&\frac{N_{\rm cr}}{S}=\frac{{4\pi }}{9}\frac{1}{{z^2 \ln ^2 z}}.
\label{ass1}
\end{eqnarray}

The asymptotics \eqref{ass1} is shown by dashed line on
Fig.~\ref{fig:w4}a. In the main logarithmic approximation this gives
simple relations
\begin{eqnarray}
&& {\mathcal E}_{1, \rm cr }=4\pi JS^2 \left(1 + C_1\sqrt
{K/J}\right),\label{ass12a} \\
&& N_{1, \rm cr } =\frac{SC_2}{\sqrt {K/J}}, \label{ass12b}
\end{eqnarray}
which coincide with the results obtained in Ref.~\onlinecite{my} by
direct numerical modeling of the discreet model, giving $C_1\approx
1.87$ and $C_2\approx 5.65$.

These expressions give us an idea about behavior of energy and bound
magnons number at small $K$. Namely we see that the energy has the
(approximate) square root singularity, while magnons number $N$
obeys inverse square root law. The asymptotics \eqref{ass12a} is
shown on Fig.~\ref{fig:w4}a by dashed line.

For $\nu=2$ the same approximation gives
\begin{eqnarray}
 &&\frac{K}{J} = \frac{{0.0066482}}{{\ln ^2 z}} + \frac{{0.0132207z}}{{\ln ^2
 z}},\nonumber \\
 &&\frac{N_{\rm cr} }{S} =  - 187.047\ln z - 108.007, \nonumber \\
 &&\frac{{\mathcal E}_{\rm cr }}{4\pi JS^2 }  = 1 + \frac{0.02762}{\ln z} +
 \frac{{0.01595}}{{\ln ^2 z}}.\label{iu1}
\end{eqnarray}
This also yields the divergent $N_{2,{\rm cr} }(K)$ and square root
peculiarity in the energy ${\mathcal E}_{2,{ \rm cr}}(K)$. However,
the coefficient before $\sqrt {K/J}$ in the energy is much smaller
then that at $\nu=1$, which makes it almost invisible in the scale
of Fig.~\ref{fig:w4}. The details of this behavior are reported on
the inset to Fig.~\ref{fig:w4}a.

This shows the similarities and differences between cases $\nu=1$
and $\nu=2$, which can also be seen from Fig.~\ref{fig:w4}. Namely,
if the topological energy densities for skyrmions with $\nu=1$ and 2
have both square-root nonanalyticity, for $\nu=2$  this
nonanalyticity reveals only as a small cusp in a close vicinity of
$K=0$. Other peculiarity of $\nu=2$ case is nonmonotonous behavior
of ${\mathcal E}_{\rm cr 2}$ (decreasing at $K<0.1J$ with subsequent
increase). Our analysis shows, that the above peculiarities of the
skyrmion with $\nu=2$ are due to interplay between exponential
asymptotics of $\theta (r)$, necessary to cut-off the divergence at
$\nu=1$ and 2 as well as power-law asymptotics, which is sufficient
for the convergence at $\nu>2$.

\section{Discussion and concluding remarks}
In this paper we present a comprehensive theoretical study of the
localized topological solitons (skyrmions), stabilized by
precessional spin dynamics, for the classical 2D ferromagnet with
easy-axial anisotropy on a square lattice. Our efforts were directed
primarily towards the study of the role of higher $\pi_2$
topological charges $\nu>1$ on the above solitons properties. Our
main conclusion is that the interplay between high topological
charges, effects of lattice discreteness (in the form of higher
powers of magnetization gradients) and uniaxial  magnetic anisotropy
makes many unexpected peculiarities into solitons properties as
compared to those in the simplest isotropic continuous model with
$W_4=0$, containing BP solitons.

Similar to previous studies, it turns out that the presence of even
weak anisotropy makes solitons dynamic, i.e., those with nonzero
precession frequency for any number $N$ of bound magnons. The
minimal consideration of discreteness (via higher degrees of
gradients of magnetization) yields the existence of some threshold
value of both soliton energy (topological energy density
\eqref{qe3}, which is more appropriate characteristic for solitons
with $\nu>1$) and the number of bound magnons. Similar to the
problem of cone state vortices,~\cite{IvWysin02} the instability is
related to the joint action of discreteness and anisotropy.
%``responsible'' for length scale $l_0$ formation (see
%Eq.~\eqref{lzero})-tut ne aysno, chto zhe est' etot scale
%u nas s Gary bylo a(J/K)^1/4, a ne (J/K)^1/2, tak chto
%luchshe ne pisat' nichego
As a result, the critical values of bound magnons $N_{\nu , \rm
{cr}}$ and soliton energy ${\mathcal E}_{\nu , \rm {cr}}$ is present
and nonanalytic dependencies of above threshold values on the
anisotropy constant appear at $\nu=1$ and $\nu=2$. It was shown
earlier,~\cite{my} that the variational minimization of $W_2$ (i.e.
energy, incorporating only squares of magnetization gradients) gives
no threshold for soliton existence, i.e. the soliton exists
everywhere up to $N=0$. This means, that mapping of the initial
discrete model even for small anisotropy $K_{\rm eff} \ll J$ on the
simplest continuum model \eqref{qe1} is wrong and to get the correct
description of solitons in 2D FM we have to consider at least fourth
powers of magnetization gradients. This seemingly paradoxical result
is actually due to the fact that the terms with $\left(\nabla {\vec
m}\right)^2$ are scale-invariant (so that the corresponding energy
has a saddle point) while the (stable) soliton size is determined by
the fourth derivatives as well as by a magnetic anisotropy. Our
analysis show, that higher (then fourth) powers of magnetization
gradients do not change the situation qualitatively, rather, in the
range of above-studied $K_{\rm {eff}}$ these terms make only a small
quantitative contribution to the solitons phase diagram. This means,
that solitons with $\nu=1$ can be well studied within the model
\eqref{kl1}.

In summary, we have shown the existence of stable topological $\pi
_2$ solitons (skyrmions) in the 2D ferromagnet with uniaxial
anisotropy on a square lattice. Since for any 2D lattice symmetry
the structure of corresponding energy functional \eqref{kl2} is the
same, we may speculate that such textures exist in any 2D magnet
with uniaxial anisotropy. We also show that due to different
character of the energy and magnons number integrals convergence at
$\nu \leq 2$ and $\nu > 2$ determines the features of so-called
solitons phase diagram. The main nontrivial and unexpected result of
the paper is that while the solitons with $\nu=1$ and $\nu >2$ have
monotonously growing phase boundary functions ${\cal E}_{\nu , {\rm
cr}}(N_{\nu , {\rm cr}})$, the case $\nu=2$ has peculiar
nonmonotonous behavior, determining the transition regime from low
to high topological charges. This means that the designated value of
soliton topological charge, which is expected for highly excited
state of FM, is neither $\nu=1$ nor high charges, but rather $\nu=2$
or $\nu=3$.

\emph{Acknowledgments.--} This work was supported  by the grant
INTAS-05-1000008-8112 and by joint grant \# 219 - 08 from the
Russian Foundation for Basic Research and Ukrainian Academy of
Science.


\begin{thebibliography}{10}
\bibitem{sn} The models with $A_\nu<0$ can be constructed, in
principle, considering different exotic types of exchange
interactions. Such consideration, however, is beyond the scope of
present paper.
\end{thebibliography}
\end{document}